\documentclass[a4paper,12pt]{article}
\usepackage{amsmath,amssymb,amsthm,epic,eepic,graphicx}
\usepackage[english]{babel}
\advance\textwidth 22mm
\advance\oddsidemargin -11mm
\advance\textheight 25mm
\advance\topmargin -15mm

\title{\LARGE\sc Q$_4$}
\author{\normalsize\sc V.E.\,Adler$^{*,a}$ and Yu.B.\,Suris$^{*,s}$}
\date{\empty}

\def\a{\alpha}
\def\b{\beta}
\def\g{\gamma} \def\G{\Gamma}
\def\d{\delta} \def\pd{\partial}
\def\eps{\varepsilon}
\def\s{\sigma}
\def\z{\zeta}

\def\Integer{\mathbb Z}
\def\Complex{\mathbb C}
\def\Real{\mathbb R}
\def\CP{\mathbb C\mathbb P}

\def\cE{{\cal E}}
\def\cL{{\cal L}}
\def\cX{{\cal X}}

\def\rL{{\rm L}}

\def\tot{\leftrightarrow}
\def\taQ#1{\label{Q#1}\tag{\text{Q$_{#1}$}}}

\def\({\left(}  \def\){\right)}
\def\<{\langle} \def\>{\rangle}

\def\const{\mathop{\rm const}}

\def\setcircle#1{\def\rrr{#1}}
\def\Black(#1,#2){\put(#1,#2){\circle*{\rrr}}}
\def\White(#1,#2){\put(#1,#2){\circle{\rrr}}}

\newtheorem{theorem}{Theorem}
\newtheorem{lemma}[theorem]{Lemma}
\newtheorem{proposition}[theorem]{Proposition}

\newbox\meibox
\def\placeunder#1#2#3#4{\setbox\meibox%
\vbox{\hbox{\hskip#4$\hphantom{#2}$}\hbox{$\hphantom{#1}$}}%
\vtop{\baselineskip=0pt\lineskiplimit=\baselineskip%
\lineskip=#3\hbox to \wd\meibox{\hfil\hskip#4$#2$\hfil}%
\hbox to \wd\meibox{\hfil$#1$\hfil}}}
\def\undertilde#1{\mathchoice{%
\placeunder{\vbox to 1.4pt{\hbox{$\displaystyle\widetilde{\,\,\,
}$}\vss}}{\displaystyle#1}{1.5pt}{1.5pt}}%
{\placeunder{\vbox to 1.4pt{\hbox{$\textstyle\widetilde{\,\,
}$}\vss}}{\textstyle#1}{1.5pt}{1.5pt}}%
{\placeunder{\vbox to 1.4pt{\hbox{$\scriptstyle\tilde{
}$}\vss}}{\scriptstyle#1}{1pt}{1pt}}%
{\placeunder{\vbox to 1.4pt{\hbox{$\scriptscriptstyle\tilde{
}$}\vss}}{\scriptscriptstyle#1}{1pt}{1pt}}%
}

\def\ti{\widetilde}
\def\tu{\undertilde}

\begin{document} \def\figurename{Fig.}
\maketitle

\let\oldf=\thefootnote \renewcommand{\thefootnote}{}
\footnotetext{$^*$ Institut f\"ur Mathematik, Technische Universit\"at Berlin,
 Str.~des 17.~Juni 136, 10623 Berlin, Germany.}
\footnotetext{$^a$ On leave from Landau Institute for Theoretical Physics,
 Chernogolovka, Russia. Supported by the Alexander von Humboldt Stiftung and
 by the RFBR grant 02-01-00144. {\tt<adler@itp.ac.ru>}}
\footnotetext{$^s$ Supported by the SFB 288 ``Differential Geometry and
 Quantum Physics''. {\tt<suris@sfb288.math.tu-berlin.de>}}
\let\thefootnote=\oldf

\section{Introduction}\label{s:intro}

One of the most fascinating and technically demanding parts of the
theory of two-dimensional integrable systems constitute the models
with the spectral parameter on an elliptic curve. These include
Landau-Lifshitz (LL) and Krichever-Novikov (KN) equations, as well
as elliptic Toda lattice (ETL) and elliptic Ruijsenaars-Toda
lattice (ERTL). The latter two systems are less known; they will
be discussed in detail in this paper (the word ``elliptic'' in
their names refers to their relation to an elliptic curve). We
will explain how all these models can be unified on the basis of a
single equation introduced in \cite{A97} as the  nonlinear
superposition formula for B\"acklund transformations of the KN
equation, and studied further in \cite{ABS,N}. We will denote it
\ref{Q4}, following \cite{ABS}. This discrete model plays the role
of the ``master equation'' in the ($sl_2$ part of) 2D
integrability: most of other models in this area can be obtained
from this one by certain limiting procedures. Some of the
interrelations have been known for a while, but several important
parts of the picture were discovered only recently. The tower of
continualizations is illustrated by the following diagram:

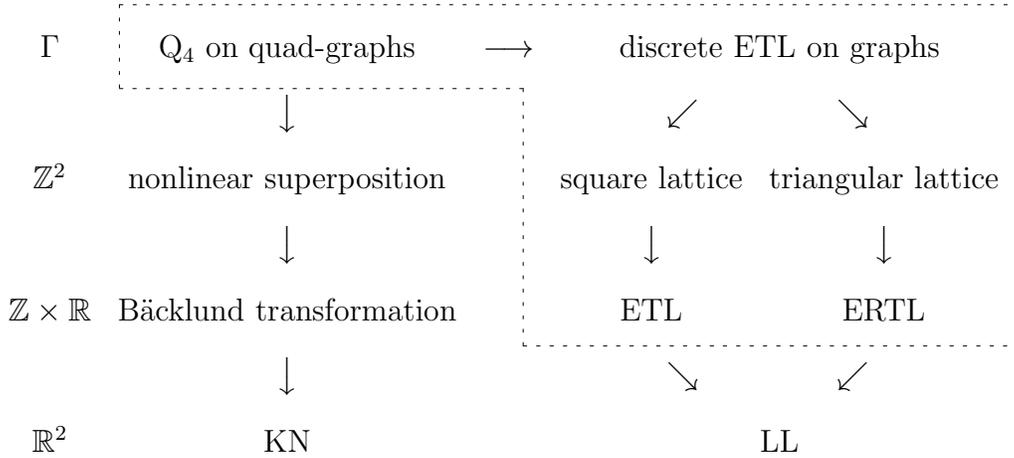
\def\vv{\vrule height1.8em width0em depth1.2em}
\setlength{\unitlength}{0.1em}
\[
 \begin{picture}(0,0)(0,0)
  \dashline{1.5}(40,80)(40,53)(170,53)(170,-30)(330,-30)(330,80)(40,80)
 \end{picture}
 \begin{array}{ccccc}
 \G & \text{\ref{Q4} on quad-graphs} &\longrightarrow &
      \multicolumn{2}{c}{\text{discrete ETL on graphs}} \vv\\
    & \big\downarrow   & & \qquad\swarrow & \searrow\qquad \\
 \Integer^2 & \text{nonlinear superposition}
      && \text{square lattice} & \text{triangular lattice} \vv\\
    & \big\downarrow & & \big\downarrow & \big\downarrow \\
 \Integer\times\Real & \text{B\"acklund transformation}
      && \text{ETL} & \text{ERTL} \vv\\
    & \big\downarrow  & & \qquad\searrow & \swarrow\qquad \\
 \Real^2 & \text{KN}
    & & \multicolumn{2}{c}{\text{LL}} \vv
 \end{array}
\]
Here the left column denotes the independent variables domain of
the system. The first line contains totally discrete systems on
planar graphs studied recently in
\cite{A00a,A01,ABS,BS02a,BHS,BH}, see also \cite{M,ND}. The link
between \ref{Q4} and discrete ETL is a sort of reduction related
with black-white coloring of quad-graphs, which was discovered in
\cite{BS02a}. The transition $\G\to\Integer^2$ is just the
specification of the general construction to the most important
cases of regular lattices. Discrete equations in this more
traditional setting were studied for a long time, see e.g.
\cite{Hir,QNCL,NC,Sur96,Sur97b,Sur97c,A00a,A00c}.

The arrows $\Integer^2\to\Integer\times\Real\to\Real^2$ have a
double meaning: a continuous system can be considered as a
(non-classical) symmetry of the discrete one, or alternatively it
can be obtained by a continuous limit. The amount of papers
growths exponentially at each step down the diagram. We mention
only the classical sources \cite{Toda,FT,Rui} and classification
results on lattice systems due to the Siberian school
\cite{Y83,SY90,Y93,AS97a}.

The contents of our paper correspond to the route along the arrows
in the framed part of the above diagram (the rest part can be
found, e.g.~in \cite{ABS,A00c}). In the main text we discuss only
the non-degenerate case related to an elliptic curve. Several more
simple cases can be obtained without additional efforts by some
specification of the general formulae, for example by substituting
$\sinh(x)$ or $x$ instead of the Weierstrass function $\s(x)$.
This corresponds to a degeneration of an elliptic curve into a
rational one. All constructions of the present paper remain valid
under such degenerations, which constitute the list
\ref{Q1}--\ref{Q4} of integrable discrete equations on quad-graphs
placed in Appendix \ref{Appendix A}. Actually, it contains six
equations, since the parameter $\d$ in equations \ref{Q1} and
\ref{Q3} can be scaled to either 0 or 1. This list represents the
most essential part of the classification result obtained in
\cite{ABS}.

\section{Discrete equations on quad-graphs}\label{s:4}

\begin{figure}[tb]
\begin{center}\setlength{\unitlength}{0.06em}
\begin{picture}(140,130)(-20,-10)
 \put(-12,110){$u_3$}            \put(  100,110){$u_2$}
 \White(0,100) \put(47,107){$\a$} \Black(100,100)
 \put(-15, 47){$\b$}              \put(  108, 47){$\b$}
 \Black(0,  0) \put(47,-15){$\a$} \White(100,  0)
 \put(-12,-15){$u_0$}            \put(  100,-15){$u_1$}
 \path(0,97)(0,0)(97,0) \path(3,100)(100,100)(100,3)
 \dashline{2}(-3,50)(103,50) \dashline{2}(-3,-3)(103,103)
 \dashline{2}(50,-3)(50,103) \dashline{2}(-3,103)(103,-3)
\end{picture}
\caption{The basic quadrilateral and $D_4$ group of symmetry}
\label{fig:quadrilateral}
\end{center}
\end{figure}
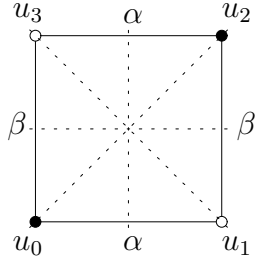

We start with equations
\begin{equation}\label{Q}
 Q(u_0,u_1,u_2,u_3;\a,\b)=0
\end{equation}
on quad-graphs (that is, planar graphs with quadrilateral faces).
Here the ``fields'' $u_k\in\CP^1$ are assigned cyclically to the
vertices of any face of the graph and the parameters
$\a,\b\in\Complex$ are assigned to its edges. It is required that
opposite edges of any face carry one and the same parameter (see
Fig.~\ref{fig:quadrilateral}). According to \cite{BS02a,N}, {\em
integrability} of such equation is synonymous with {\em 3D
consistency}, see below. The most important representative of
integrable equations (\ref{Q}) is the equation \ref{Q4}, which we
give here in the nice form derived in \cite{N}:
\begin{alignat}{1}
  & A\big((u_0-b)(u_3-b)-(a-b)(c-b)\big)
     \big((u_1-b)(u_2-b)-(a-b)(c-b)\big) \nonumber\\
 +& B\big((u_0-a)(u_1-a)-(b-a)(c-a)\big)
     \big((u_3-a)(u_2-a)-(b-a)(c-a)\big) \nonumber\\
  & \quad =ABC(a-b),\taQ4
\end{alignat}
where the points
\[
 (a,A)=(\wp(\a),\wp'(\a)),\quad (b,B)=(\wp(\b),\wp'(\b)),\quad
 (c,C)=(\wp(\b-\a),\wp'(\b-\a))
\]
belong to the elliptic curve $\cE=\{A^2=r(a)\}$, and
$r(a)=4a^3-g_2a-g_3$ is the notation for the Weierstrass
polynomial used throughout the paper. So, in this case it is
natural to assume that the parameters $\a,\b$  belong to the
period parallelogram of $\cE$ rather than to $\Complex$. A
collection of formulas related to $\cE$ and to the corresponding
Weierstrass elliptic functions is put in Appendix \ref{Appendix
B}.

It is obvious that \ref{Q4} is invariant under reflections in the
diagonals: $u_0\tot u_2$, $\a\tot\b$ and $u_1\tot u_3$,
$\a\tot\b$, as well as under reflections $(u_0,u_3)\tot(u_1,u_2)$
and $(u_0,u_1)\tot(u_3,u_2)$, so that this equation admits the
symmetry group $D_4$ of the square.

\setcircle{6}

Actually, the symmetry of \ref{Q4} is even more rich. This will follow from
another form of this equation, which is of a fundamental importance also for
many other reasons, as will be shown below.

\begin{proposition}\label{prop:3leg}
Equation \ref{Q4} is equivalent, under the point transformations
$u_i=\wp(x_i)$, to equation
\begin{equation}\label{3leg mult}
  F(x_0,x_1;\a)/F(x_0,x_3;\b)=F(x_0,x_2;\a-\b),
\end{equation}
where
\begin{equation}\label{Q4 F}
 F(x_0,x_1;\a)=\frac{\s(x_0+x_1+\a)\s(x_0-x_1+\a)}
                    {\s(x_0+x_1-\a)\s(x_0-x_1-\a)}.
\end{equation}
\end{proposition}

This proposition was proved in \cite{ABS}. Equation (\ref{3leg
mult}) is called the {\em multiplicative three-leg form} of
(\ref{Q}) {\em centered at $x_0$}. Sometimes it is more convenient
to use the {\em additive three-leg form}
\begin{equation}\label{3leg add}
 f(x_0,x_1;\a)-f(x_0,x_3;\b)=f(x_0,x_2;\a-\b),
\end{equation}
where
\begin{equation}\label{f}
f(x_0,x_1;\a)=\frac12\log F(x_0,x_1;\a).
\end{equation}
The functions $f$ for equations \ref{Q1}--\ref{Q3}, as well as the
corresponding point transformations of the fields and parameters
are listed in Appendix \ref{Appendix A}.

Proposition \ref{prop:3leg} is checked with the help of some
identities for elliptic functions listed in Appendix \ref{Appendix B}. 
With the help of (\ref{p-p}) one sees immediately that the
function (\ref{Q4 F}) is fractional-linear in $u_1=\wp(x_1)$ (but
not in $u_0=\wp(x_0)$):
\begin{equation}\label{Ffl}
 F(x_0,x_1;\a)=
 \frac{\wp(x_1)-\wp(x_0+\a)}{\wp(x_1)-\wp(x_0-\a)}
 \cdot\frac{\s^2(x_0+\a)}{\s^2(x_0-\a)}.
\end{equation}
Therefore, the formula (\ref{3leg mult}) yields an equation which
is already affine-linear in $u_1,u_2,u_3$, but with a complicated
dependence on $x_0$. Some additional transformations based on 
(\ref{4sigmas}), (\ref{det to sigma}) are needed in
order to show that this equation coincides with \ref{Q4} up to
some factor depending on $x_0$. Note that eq. (\ref{3leg mult})
defines, strictly speaking, a correspondence, since each $x_i$ is
defined up to a sign and up to shifts by periods of the function
$\wp(x)$.

The function $F$ has an obvious symmetry property
\begin{equation}\label{Fsym}
 F(x_0,x_1;\a)=\frac1{F(x_0,x_1;-\a)}
\end{equation}
which implies one of the $D_4$ symmetries $x_1\tot x_3,$
$\a\tot\b$. However, another symmetry $(x_0,x_3)\tot (x_1,x_2)$ of
the equation \ref{Q4} becomes hidden in its three-leg form
(\ref{3leg mult}) centered at $x_0$. {\em Due to this symmetry the
three-leg form can be centered at an arbitrary vertex.}

On the other hand, three-leg form exhibits one more symmetry which
remains hidden in the rational form \ref{Q4}, namely the fact that
the diagonals of the quadrilateral are on the equal footing with
its edges. Indeed, denote $\a=\a_1-\a_2$, $\b=\a_3-\a_2$, then eq.
(\ref{3leg mult}) takes the form
\begin{equation}\label{3leg sym}
 F(x_0,x_1;\a_1-\a_2)F(x_0,x_2;\a_2-\a_3)F(x_0,x_3;\a_3-\a_1)=1.
\end{equation}
This means that the full symmetry group of discrete KN equation is
not the $D_4$ group of the square on Fig.~\ref{fig:quadrilateral},
but the symmetric group $S_4$ of the tetrahedron on
Fig.~\ref{fig:tetr}. Probably, there does not exist the form of
equation, in which the whole group becomes apparent. Indeed, even
in the simplest case of equation (\ref{Q1})$_{\d=0}$ (see Appendix
\ref{Appendix A}) which reads  {\sl
CrossRatio}$\,(x_0,x_1,x_2,x_3)=\a/\b$, this full group $S_4$ is
encoded in the transformation properties of the cross-ratio.

\setcircle{5}
\begin{figure}[tb]
\begin{center}\setlength{\unitlength}{0.08em}
\begin{minipage}[t]{70mm}
\begin{picture}(160,150)(-55,-35)
 \Black(35,-25)\Black(-35,25)\Black(45,120)\Black(120,0)
 \path(35,-25)(-35,25)(45,120)(120,0)(35,-25)(45,120)
 \dashline{4}(-35,25)(120,0)
 \put(50,120){$x_0$}
 \put(-51,20){$x_1$}\put(25,-34){$x_2$}\put(115,-12){$x_3$}
 \put(-40,74){$\a_1-\a_2$}\put(42,30){$\a_2-\a_3$}\put(88,54){$\a_3-\a_2$}
\end{picture}
\caption{$S_4$ symmetry of discrete KN}
\label{fig:tetr}
\end{minipage}
\quad 
\begin{minipage}[t]{70mm}
\begin{picture}(170,180)(-35,-10)
 \path(0,120)(60,160)(140,140)(140,40)(80,0)(0,20)(0,120)(80,100)(140,140)
 \path(80,0)(80,100)
 \dashline{6}(0,20)(60,60)(140,40)\dashline{6}(60,60)(60,160)
 \Black(0,20)\Black(60,160)\Black(80,100)\Black(140,40)
 \dashline{3}(0,20)(60,160)(140,40)(0,20)(80,100)(60,160)
 \dashline{3}(80,100)(140,40)
 \put(79,-11){0}\put(145,36){1}\put(-11,16){2}\put(55,46){12}
 \put(68,105){3}\put(144,136){31}\put(-16,116){23}\put(50,166){123}
 \put(111,12){$\a_1$}\put(33,-2){$\a_2$}\put(-14,66){$\a_3$}
\end{picture}
\caption{3D consistency}
\label{fig:3d}
\end{minipage}
\end{center}
\end{figure}

As the first application of the three-leg form, we will prove that
it (together with the above mentioned symmetry properties) implies
the fundamental property of 3D consistency, which can be accepted
as the definition of integrability for equations of the form
(\ref{Q}). Recall \cite{BS02a,N} that equation (\ref{Q}) is called
3D-consistent, if the following holds. For arbitrary initial data
$u_0,u_1,u_2,u_3$, define the values $u_{12},u_{23},u_{31}$ (the
enumeration, which is different from the one used elsewhere in
this paper, is explained on Fig.~\ref{fig:3d}) according to the
equations
\begin{align}
 Q(u_0,u_1,u_{12},u_2;\a_1,\a_2)&=0, \label{Q12}\\
 Q(u_0,u_2,u_{23},u_3;\a_2,\a_3)&=0, \label{Q23}\\
 Q(u_0,u_3,u_{31},u_1;\a_3,\a_1)&=0, \label{Q31}
\end{align}
which correspond to three faces of the cube adjacent to the vertex 0.
Then equations
\begin{align}
 Q(u_1,u_{12},u_{123},u_{31};\a_2,\a_3)&=0, \label{Q123}\\
 Q(u_2,u_{23},u_{123},u_{12};\a_3,\a_1)&=0, \label{Q231}\\
 Q(u_3,u_{31},u_{123},u_{23};\a_1,\a_2)&=0, \label{Q312}
\end{align}
corresponding to the rest faces, define one and the same value $u_{123}$.

\begin{proposition}\label{prop:3d}
Suppose that equation (\ref{Q}) admits the $D_4$ symmetry group and is
equivalent, under some point transformation $u=\phi(x)$, $a=\rho(\a)$, to
the three-leg equation (\ref{3leg mult}). Then it is 3D-consistent.
\end{proposition}
\begin{proof}
It is enough to show that if $u_{123}$ is defined by equation
(\ref{Q123}), then (\ref{Q231}) is fulfilled as well. Rewrite equations
(\ref{Q12}), (\ref{Q31}), (\ref{Q123}) (corresponding to the faces
adjacent to the vertex 1) in the three-leg forms centered at $x_1$ (recall
that this is possible due to the symmetry properties):
\begin{align*}
  F(x_1,x_{31};\a_3)/F(x_1,x_0;\a_1)   &= F(x_1,x_3;\a_3-\a_1),\\
  F(x_1,x_{12};\a_2)/F(x_1,x_0;\a_1)   &= F(x_1,x_2;\a_2-\a_1),\\
  F(x_1,x_{12};\a_2)/F(x_1,x_{31};\a_3)&= F(x_1,x_{123};\a_2-\a_3).
\end{align*}
From these there follows the equation relating the fields at the vertices
of the dashed tetrahedron on Fig.~\ref{fig:3d}:
\[
  F(x_1,x_2;\a_2-\a_1)/F(x_1,x_3;\a_3-\a_1)=F(x_1,x_{123};\a_2-\a_3).
\]
This is nothing but the three-leg form of the equation
\begin{equation}\label{tetr}
 Q(u_1,u_2,u_3,u_{123};\a_2-\a_1,\a_2-\a_3)=0,
\end{equation}
centered at $x_1$. This can be centered at $x_2$, as well, resulting in
the cyclic shift of indices:
\[
  F(x_2,x_3;\a_3-\a_2)/F(x_2,x_1;\a_1-\a_2)=F(x_2,x_{123};\a_3-\a_1).
\]
The latter equation, together with the three-leg forms of equations
(\ref{Q12}), (\ref{Q23}) centered at $x_2$, leads to the three-leg form of
(\ref{Q231}), as required.
\end{proof}

\section{Discrete Toda systems on graphs}\label{sect: dToda}

Existence of the three-leg form allows us to establish a direct
link to discrete Toda systems on graphs introduced in \cite{A01}.
This link was discovered in \cite{BS02a} and is described as
follows. Suppose that the quad-graph $\G$ is bi-partite, i.e. its
vertices can be colored black and white so that the ends of any
edge are of different colors. Consider the graph $\G_0$ whose set
of vertices $V(\G_0)$ consists of the black vertices of $\G$, and
whose edges are diagonals of quadrilateral faces of $\G$. Consider
a vertex $x_0\in V(\G_0)$ (for brevity we will identify vertices
with the field variables sitting there), and let $x_0$ be a common
vertex of $n$ adjacent quadrilateral faces
$(x_0,x_{2j-1},x_{2j},x_{2j+1})\in F(\Gamma)$, $j=1,\ldots,n$,
like shown on Fig.~\ref{fig:flower}. Then, multiplying the
equations of the type (\ref{3leg mult}) centered in $x_0$ for all
these faces, we see that the white vertices $x_{2j-1}$ cancel out
and the black vertices $x_{2j}$ satisfy the equation on the star
of the vertex $x_0$:
\begin{equation}\label{mToda}
 \prod_{j=1}^nF(x_0,x_{2j};\a_j-\a_{j+1})=1,
\end{equation}
or, in the additive form,
\begin{equation}\label{Toda}
 \sum_{j=1}^nf(x_0,x_{2j};\a_j-\a_{j+1})=0.
\end{equation}
This is a {\em discrete Toda system on the graph $\G_0$}. (Of
course, a similar Toda type system holds also for the white graph
$\G_1$ which is dual to $\G_0$.) The parameters $\a_j$ which were
assigned originally to the edges $(x_0,x_{2j-1})\in E(\G)$, label
now the corners of the faces of $\G_0$. Notice that only the {\em
differences} of these parameters appear in equations (\ref{Toda}),
so that simultaneous shifting all the parameters of the equations
on the original quad-graph $\a\to\a+\lambda$ leads to the same Toda
system on $\G_0$. This shift plays the role of the spectral
parameter in the zero curvature representation of the Toda system.

As established in \cite{BS02a, N}, any 3D consistent equation
(\ref{Q}) is its own zero curvature representation. We will show
that it plays this role not only for itself, but for a whole set
of related systems, including discrete Toda systems.
\setcircle{5}
\begin{figure}[tb]\setlength{\unitlength}{0.08em}
\def\Bray(#1,#2){\Black(#1,#2)}
\def\tmpg{\Bray(0,0)
  \Bray(70,30)  \put(76,30){$x_2$}
  \Bray(10,80)  \put(15,80){$x_4$}
  \Bray(-70,-10)\put(-83,-17){$x_6$}
  \Bray(0,-70)  \put(5,-73){$x_8$}
  \Bray(80,-30) \put(85,-30){$x_{10}$}}
\begin{center}
\begin{minipage}[t]{70mm}
\begin{picture}(200,160)(-100,-80)
 \tmpg \put( -5,-11){$x_0$}
 \path(0,0)(  27.5 ,  0   )\White( 30,  0) \put( 38, -2){$x_1$}
 \path(0,0)(  19   , 38   )\White( 20, 40) \put( 25, 45){$x_3$}
 \path(0,0)( -28.35, 28.35)\White(-30, 30) \put(-46, 31){$x_5$}
 \path(0,0)( -38.3 ,-28.5 )\White(-40,-30) \put(-54,-37){$x_7$}
 \path(0,0)(  28.35,-28.35)\White( 30,-30) \put( 33,-39){$x_9$}
 \path(31.8,1.7)(70,30)(22.4,39)
 \path(19,42)(10,80)(-28.5, 31.7)
 \path(-32,28.8)(-70,-10)(-41.8,-28.6)
 \path(-38.3,-31.7)(0,-70)(28.6,-31.8)
 \path(32.5,-30)(80,-30)(31.8,-1.7)
 \put( 13,  3){$\a_1$}
 \put( -3, 22){$\a_2$}
 \put(-25,  8){$\a_3$}
 \put(-25,-25){$\a_4$}
 \put( 17,-15){$\a_5$}
 \end{picture}
\caption{Faces adjacent to the vertex $x_0$.}\label{fig:flower}
\end{minipage}
\quad 
\begin{minipage}[t]{70mm}
\def\Bray(#1,#2){\path(0,0)(#1,#2)\Black(#1,#2)}
\begin{picture}(200,160)(-100,-80)
 \tmpg
 \put( 10, -1){$\a_1$} \put(52,30){$\a_1$}
 \put(  3, 10){$\a_2$} \put(60,18){$\a_2$}
 \put(-15,  5){$\a_3$}
 \put(-14,-10){$\a_4$}
 \put(  5,-13){$\a_5$}
\end{picture}
\caption{The black graph star of the vertex $x_0$.}
\label{fig:star}
\end{minipage}
\end{center}
\end{figure}
For this aim, rewrite eq. \ref{Q4} on the quadrilateral
$(u_0,u_1,u_2,u_3)\in F(\G)$ as
\begin{multline*}
  k_0u_0u_1u_2u_3+k_1(u_1u_2u_3+u_0u_2u_3+u_0u_1u_3+u_0u_1u_2)
  +k_2(u_0u_2+u_1u_3) \\
  +k_3(u_0u_1+u_2u_3)+k_4(u_0u_3+u_1u_2)+k_5(u_0+u_1+u_2+u_3)+k_6=0.
\end{multline*}
(Concrete expressions of the coefficients $k_i$ in terms of
$(a,A)$ and $(b,B)$ can be found in the original paper
\cite{A97}.) The field $u_3$ is a fractional-linear function
(M\"obius transformation) of $u_1$ with coefficients depending on
$u_0,u_2$ and $\a_1,\a_2$. Denote the matrix of this M\"obius
transformation as
\begin{equation}\label{U}
 U(u_0,u_2;\a_1,\a_2)=\frac1D
 \begin{pmatrix}
   -k_1u_0u_2-k_3u_0-k_4u_2-k_5 & -k_2u_0u_2-k_5(u_0+u_2)-k_6\\
    k_0u_0u_2+k_1(u_0+u_2)+k_2  & k_1u_0u_2+k_4u_0+k_3u_2+k_5
 \end{pmatrix},
\end{equation}
where $D$ is the normalizing factor such that $\det U=1$.
Considering now $n$ quadrilateral faces
$(u_0,u_{2j-1},u_{2j},u_{2j+1})\in F(\Gamma)$ adjacent to $u_0$,
we see that, according to the cancellation property leading to the
Toda systems, the product
\[
 \prod_{1\le j\le n}^\curvearrowleft U(u_0,u_{2j};\a_j,\a_{j+1})
\]
generates an identical M\"obius transformation and is therefore a scalar
matrix. Moreover, the latter property is equivalent to the discrete Toda
equation at the vertex $u_0\in V(\G_0)$, formulated now in terms of the
variables $(u_0,u_{2j})$. Thus, we see that the following holds.

\begin{proposition}\label{prop: dToda ZCR}
Discrete Toda equation (\ref{mToda}) at the vertex $u_0\in V(\G_0)$ is
rational in terms of $u_0$ and $u_{2j}$, $j=1,\ldots,n$. The discrete Toda
system on an arbitrary graph $\G_0$ admits a zero curvature representation
in the sense of \cite{A01}, with the transition matrices across the edge
$(u_0,u_2)\in E(\G_0)$ given by $U(u_0,u_2;\a_1+\lambda,\a_2+\lambda)$.
\end{proposition}

The fact that the Toda equation for the star of the vertex $u_0$
is rational with respect to $u_{2j}=\wp(x_{2j})$, $j=1,\ldots,n$,
is easy to see from (\ref{mToda}), which can be rewritten with the
help of the formula (\ref{Ffl}) as
\[
 \prod^n_{j=1}\frac{\wp(x_{2j})-\wp(x_0+\a_j-\a_{j+1})}
 {\wp(x_{2j})-\wp(x_0-\a_j+\a_{j+1})}
 \cdot \frac{\s^2(x_0+\a_j-\a_{j+1})}{\s^2(x_0-\a_j+\a_{j+1})}=1.
\]
However, the rationality with respect to $u_0=\wp(x_0)$ remains
hidden in this representation. This is a new manifestation of the
fact that the original (polynomial in $u_j$) and the three-leg
forms of the equation \ref{Q4} are complementary in showing and
hiding various properties.

The next, Lagrangian property is better visible again in the
variables $x$, with the help  of the three-leg form. It is easy to
see that the function $\pd f(x_0,x_1;\a)/\pd x_1$ is symmetric
with respect to the flip $x_0\tot x_1$. Integrating this function
with respect to $x_0$ and $x_1$, we see that there exists a
symmetric function $\Lambda(x_0,x_1;\a)=\Lambda(x_1,x_0;\a)$, such
that
\[
 f(x_0,x_1;\a)=\pd\Lambda(x_0,x_1;\a)/\pd x_0.
\]

\begin{proposition}\label{prop:EL}
Let $E(\G_0)$ be the set of the edges of the black graph $\G_0$.
Let the pairs of parameters be assigned to the edges from
$E(\G_0)$ according to Fig.~\ref{fig:star}, so that, e.g., the
pair $(\a_1,\a_2)$ corresponds to the edge $(x_0,x_2)$. Then the
discrete Toda equations (\ref{Toda}) are Euler--Lagrange equations
for the action functional
\[
 S=\sum_{(x_0,x_2)\in E(\G_0)}\Lambda(x_0,x_2;\a_1-\a_2).
\]
\end{proposition}

Finally, we give the matrix formulation for the cancellation phenomenon
which is the main feature of the three-leg form.

\begin{lemma}\label{ZCR matrices}
If the relation
\begin{equation}\label{for zcr}
 f(x_0,x_1;\a)-f(x_0,x_3;\b)=q\quad \Leftrightarrow\quad
 F(x_0,x_1;\a)/F(x_0,x_3;\b)=e^{2q}
\end{equation}
holds for some quantity $q$, then $u_3=\wp(x_3)$ is a M\"obius
transformation of $u_1=\wp(x_1)$ with the matrix
\begin{equation}\label{zcr L}
  L=e^qN(x_0;\b,\a)-e^{-q}N(x_0;-\b,-\a),
\end{equation}
where
\begin{equation}\label{zcr M}
 N(x;\b,\a)=\frac{\s^2(x+\b)\s^2(x-\a)}{\s(2x)}
  \begin{pmatrix}
   -\wp(x+\b) & \wp(x+\b)\wp(x-\a) \\ -1 & \wp(x-\a)
  \end{pmatrix}.
\end{equation}
\end{lemma}
\begin{proof}
According to (\ref{Ffl}), the formula (\ref{for zcr}) is
equivalent to:
\[
 \frac{\wp(x_3)-\wp(x_0+\b)}{\wp(x_3)-\wp(x_0-\b)}\cdot
 \frac{\s^2(x_0+\b)}{\s^2(x_0-\b)}=
 \frac{\wp(x_1)-\wp(x_0+\a)}{\wp(x_1)-\wp(x_0-\a)}\cdot
 \frac{\s^2(x_0+\a)}{\s^2(x_0-\a)}\cdot e^{-2q}.
\]
Now it is a matter of a straightforward computation to represent
this as the M\"obius transformation from $\wp(x_1)$ to $\wp(x_3)$
with the matrix $L$ given by (\ref{zcr L}), (\ref{zcr M}).
\end{proof}

Of course, matrix $L$ is defined only up to a scalar factor, which
is chosen in (\ref{zcr L}), (\ref{zcr M}) so that $\det
L=\s(2\a)\s(2\b)=\const$. The telescopic cancellation taking place
by adding several equations of the type (\ref{for zcr}), is
translated for the matrices $L$ into the following property: if,
additionally to (\ref{zcr L}), we have
\[
 L_1=e^{q_1}N(x_0;\g,\b)-e^{-q_1}N(x_0;-\g,-\b),
\]
then
\[
 L_1L=\s(2\b)\big(e^{q_1+q}N(x_0;\g,\a)-e^{-q_1-q}N(x_0;-\g,-\a)\big).
\]
This follows easily from the properties of the rank 1 matrices $N$, namely:
\[
 N(x;\g,\b)N(x;\b,\a)=\s(2\b)N(x;\g,\a),\quad
 N(x;\g,\b)N(x;-\b,-\a)=0.
\]

Notice that the case $q=f(x_0,x_2;\a-\b)$ in eq. (\ref{for zcr})
corresponds to the additive three-leg form (\ref{3leg add}) of
equation \ref{Q4}. In this case we obtain from Lemma \ref{ZCR
matrices} a new formula for the transition matrices from
Proposition \ref{prop: dToda ZCR}:
\[
 U(u_0,u_2;\a_1,\a_2)\sim  e^{f(x_0,x_2;\a_1-\a_2)}N(x_0;\a_2,\a_1)-
  e^{-f(x_0,x_2;\a_1-\a_2)}N(x_0;-\a_2,-\a_1),
\]
where symbol $\sim$ means equality up to a constant scalar factor.
A direct verification of this formula is actually equivalent to a
proof of Proposition \ref{prop:3leg}.

\section{Lagrangian systems with discrete time \\ and space-time
slicing of regular lattices}

Now we consider the specification of the general scheme
corresponding to the case of the shift-invariant graphs, or
lattices. Recall the general construction of the discrete time
Lagrangian mechanics \cite{MV}. Let $\cX$ be a vector space (the
theory can be developed also in a more general setting, when $\cX$
is a manifold, but we will not need this here). Let
$\cL:\cX\times\cX\mapsto\Real$ be a smooth function, called the
{\em discrete time Lagrange function}. For an arbitrary sequence
$X:\eps\Integer\supset\{a\eps,a\eps+\eps,\ldots,b\eps\}\mapsto\cX$
one considers the {\em discrete time action functional}
\[
 S=\sum_{n=a}^{b-1}\cL(X(n\eps),X(n\eps+\eps)).
\]
The sequences serving as critical points of the action functional
(in the class of variations preserving $X(a\eps)$ and $X(b\eps)$)
satisfy the {\em discrete time Euler-Lagrange equation}, which we
shall write in the index-free form as
\begin{equation}\label{dEL}
 \pd\cL(X,\ti X)/\pd X+\pd\cL(\tu X,X)/\pd X=0
\end{equation}
where $\tu X$, $X$, $\ti X$ stand for $X(n\eps-\eps)$, $X(n\eps)$,
$X(n\eps+\eps)$,
respectively. This is an implicit equation for $\ti X$. In general, it has
more than one solution, and therefore defines a correspondence
(multi-valued map) $(\tu X,X)\mapsto(X,\ti X)$. To discuss symplectic
properties of this correspondence, define
\[
  P=\pd\cL(\tu X,X)/\pd X\in T^*_X\cX.
\]
Then (\ref{dEL}) may be rewritten as the system
\begin{equation}\label{dEL Ham}
 P=-\pd\cL(X,\ti X)/\pd X,\quad \ti P=\pd\cL(X,\ti X)/\pd\ti X.
\end{equation}
This system will be called the {\em Hamiltonian form} of eq.
(\ref{dEL}). It defines a multi-valued map $(X,P)\mapsto(\ti X,\ti
P)$ on $T^*\cX$. More precisely, the first equation in (\ref{dEL
Ham}) is an implicit equation for $\ti X$, while the second one
allows one to calculate $\ti P$ explicitly and uniquely, once $X$
and $\ti X$ are known. The fundamental property of this
multi-valued map is the following: {\em each branch of the map
$T^*\cX\mapsto T^*\cX$ defined by (\ref{dEL Ham}) is symplectic
with respect to the standard symplectic structure on $T^*\cX$.}

The continuous limit of this construction is as follows. Suppose
that $X(n\eps)$ approximates the smooth function $X(t)$ at
$t=n\eps$ with some small $\eps>0$. Suppose that by the
substitution $(X,\ti X)=(X,X+\eps\dot X+o(\eps))$ there holds an
asymptotic relation:
\[
 \cL(X,\ti X)\approx \eps\rL(X,\dot X)
\]
with some smooth function $\rL:T\cX\mapsto\Real$, where the sign
$\approx$ means equality up to additive terms $o(\eps)$. Then we
have:
\begin{align*}
\pd\rL/\pd\dot X & \approx \pd\cL(X,\ti X)/\pd\ti X=\ti P,\\
 \pd\rL/\pd X  & \approx
  \eps^{-1}\big(\pd\cL(X,\ti X)/\pd X
     +\pd\cL(X,\ti X)/\pd\ti X\big)= \eps^{-1}(\ti P-P).
\end{align*}
Thus, the discrete Lagrangian equations (\ref{dEL Ham}) approximate (serve
as a discretization of) the continuous Euler--Lagrange equations with the
Lagrangian $\rL(X,\dot X)$:
\[
 P=\pd\rL(X,\dot X)/\pd\dot X,\quad \dot P=\pd\rL(X,\dot X)/\pd X.
\]

Now we can introduce the discrete time Lagrangian formalism for discrete
Toda systems on some regular two-dimensional lattices. We will consider
planar graphs $\G$ with the set of vertices $V(\G)=\Integer^2$ only. They
will be distinguished by their sets of edges $E(\G)$. It will be
convenient to think of the vertices $(k,n)\in\Integer^2$ as representing
the points $(k,n\eps)\in\Real^2$ with some (small) $\eps>0$. The first
index $k$ will enumerate the lattice sites, while the second index $n$
has the meaning of the discrete time $t=n\eps$. For a fixed $t=n\eps$,
we will write $V^t=\{(k,t):k\in\Integer\}$, so that
$V(\G)=\bigcup_{t\in \eps\Integer}V^t$.

A {\em slicing} of a graph $\G$ consists in choosing subgraphs
$\G^t$ so that $V(\G^t)=V^{t-\eps}\cup V^t$, the sets
$E(\G^t)=E^t$ are disjoint, and $E(\G)=\bigcup_{t\in
\eps\Integer}E^t$.

The configuration space of our discrete time systems is
$\cX=\{X:\Integer\mapsto\Real\}$ consisting of real-valued sequences
$X=(x_k)_{k\in\Integer}$. An element $P\in T^*_X\cX$ is a sequence
$P=(p_k)_{k\in\Integer}$. Clearly, for any fixed $t=n\eps$, the set of
functions on $V^t$ is identified with $\cX$. Functions on $V(\G^t)$ are
naturally identified with pairs of functions $(\tu X,X)$ on $V^{t-\eps}$
and $V^t$, respectively, and therefore belong to $\cX\times\cX$. The
discrete time Lagrange functions for our Toda systems are of the form
\[
 \cL(\tu X,X)=\sum_{(x,x')\in E^t}\Lambda(x,x';\a-\b),
\]
i.e. are obtained by summing up the elementary Lagrangians over the edges
of one slice of the graph $\G$.

The pictures of the slicing for the graphs playing the main roles in our
subsequent presentation are given below. Starting from this point, we use
the following abbreviated notation for the sequences like
$(x_k)_{k\in\Integer}$: we write $x$, $x_{\pm1}$ for $x_k$, $x_{k\pm1}$,
respectively. This notation has not to be confused with indices used in
the previous section.

\setcircle{10}
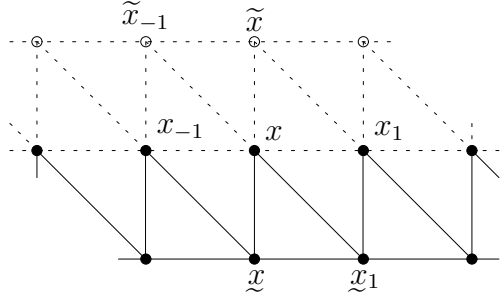
\begin{figure}[t]
\begin{center}
\setlength{\unitlength}{0.035em}
\begin{picture}(450,270)(-25,-30)
 \dashline{4}(-25,200)(325,200)
 \multiput(0,0)(100,0){4}{
  \White(0,200)\dashline{4}(0,100)(0,200)(100,100)
  \Black(100,0)\path(0,100)(100,0)(100,100)}
 \dashline{4}(-25,100)(425,100)
 \dashline{4}(-25,125)(0,100)
 \dashline{4}(400,100)(400,125)
 \multiput(0,100)(100,0){5}{\Black(0,0)}
 \path(0,75)(0,100) \path(75,0)(425,0) \path(400,100)(425,75)
 \multiputlist(100,210)(100,0)[cb]{$\ti x_{-1}$,$\ti x$}
 \multiputlist(110,110)(100,0)[lb]{$x_{-1}$,$x$,$x_1$}
 \multiputlist(200,-10)(100,0)[ct]{$\tu x$,$\tu x_1$}
\end{picture}
\caption{Slice $(4+2)$ on triangular lattice}
\label{fig:slice24}
\end{center}
\end{figure}

The central role in what follows will play the discrete Toda
system on the triangular lattice and on the square lattice. All
vertices of the regular triangular lattice have valence 6, so that
the corresponding Toda system is a 7-point scheme. 
Similarly, the Toda system on the regular square
lattice is a 5-point scheme. The Hamiltonian properties of these
systems depend essentially on the slicing chosen. For the
triangular lattice we will choose the slicing on
Fig.~\ref{fig:slice24}. The corresponding map $(\tu X,X)
\mapsto(X,\ti X)$ is defined implicitly, since each equation
contains two fields $\ti x$ and $\ti x_{-1}$ from the time level
$t=n\eps+\eps$). For the regular square lattice we will use two
different slicings. The first one is shown on
Fig.~\ref{fig:slice22} in a somewhat unusual way (by a standard
drawing the slice would run diagonally). It corresponds to an
implicit discrete Toda system. Indeed, also in this case each
equation contains two fields $\ti x$ and $\ti x_{-1}$ from the
time level $t=n\eps+\eps$. Finally, the slicing shown on
Fig.~\ref{fig:slice13} represents an explicit discrete Toda
system, since each equation contains only one updated field $\ti
x$. These two types of 5-point equations can be
obtained from the 7-point one by reduction consisting in erasing
some edges of the triangular lattice.

\begin{figure}[h]
\begin{center}\setlength{\unitlength}{0.035em}
\begin{minipage}[t]{70mm}
\begin{picture}(450,270)(-35,-30)
 \multiput(0,0)(100,0){4}{
  \White(0,200)\dashline{4}(0,100)(0,200)(100,100)
  \Black(100,0)\path(0,100)(100,0)(100,100)}
 \multiput(0,100)(100,0){5}{\Black(0,0)}
 \dashline{4}(0,100)(-25,125) \dashline{4}(400,100)(400,125)
 \path(0,75)(0,100) \path(400,100)(425,75)
 \multiputlist(100,210)(100,0)[cb]{$\ti x_{-1}$,$\ti x$}
 \put(209,100){$x$}
 \multiputlist(200,-10)(100,0)[ct]{$\tu x$,$\tu x_1$}
\end{picture}
\caption{Slice $(2+2)$ on square lattice} \label{fig:slice22}
\end{minipage}
\quad 
\begin{minipage}[t]{70mm}
\begin{picture}(450,270)(-35,-30)
 \multiput(0,0)(100,0){5}{
   \White(0,200)\dashline{4}(0,100)(0,200)
   \Black(0,100)\Black(0,0)\path(0,0)(0,100)}
 \dashline{4}(-20,200)(420,200)
 \dashline{4}(-20,100)(420,100)
 \path(-20,0)(420,0)
 \put(195,215){$\ti x$}
 \multiputlist(105,105)(100,0)[lb]{$x_{-1}$,$x$,$x_1$}
 \put(190,-20){$\tu x$}
\end{picture}
\caption{Slice $(3+1)$ on square lattice} \label{fig:slice13}
\end{minipage}
\end{center}
\end{figure}

\section{Triangular lattice}\label{s:7TL}

Time discretizations of lattices of the Ruijsenaars--Toda type
\cite{Rui} were introduced in \cite{Sur96,Sur97b}. They were
identified in \cite{A00a} as discrete systems on the regular
triangular lattice.

The regular triangular lattice $T$ can be considered as the black
subgraph of the quad-graph known as the {\em dual kagome lattice}
(drawn on Fig.~\ref{fig:L3} in dashed lines). The latter graph has
vertices of two kinds, black vertices of valence 6 and white
vertices of valence 3, and edges of three types, all edges of each
type being parallel. The parameters corresponding to these three
types of edges are denoted $\a_1,\a_2,\a_3$ as shown on
Fig.~\ref{fig:L3}. It is easy to see that the parameters $\a_3$
are constant along horizontal stripes, that is $\a_3=\a_{3,n}$ and
analogously, $\a_2=\a_{2,k}$ and $\a_1=\a_{1,k+n}$.

In what follows, we restrict ourselves to the particular case which admits
a well-defined continuous limit, when the parameters $\a_1$ and $\a_3$ are
constant:
\begin{equation}\label{dual kagome param}
 \a_1=\lambda,\quad \a_{2,k}=\lambda-\b_k,\quad \a_3=\lambda-\eps,
\end{equation}
where $\lambda$ will play the role of the spectral  parameter in the
zero curvature representation of the Toda system. The discrete
Lagrange function obtained by summing up the elementary
Lagrangians along all edges of a time slice, is equal to
\begin{equation}\label{triang Toda Lagr}
 \cL(\tu X,X)=\sum_k\bigl(
   \Lambda(x_k,\tu x_k;\eps)
  +\Lambda(x_k,\tu x_{k+1};\b_k-\eps)
  +\Lambda(\tu x_k,\tu x_{k+1};-\b_k)\bigr).
\end{equation}
\setcircle{5}
\def\tmpga(#1,#2){\put(#1,#2){\path(0,0)(0,100)(100,0)(0,0)
 \Black(0,0) \Black(100,0) \Black(0,100) \White(34,34)
 \dashline{4}(0,0)(30,30) \dashline{4}(0,100)(32,39)
 \dashline{4}(100,0)(39,32)}}
\def\tmpgb(#1,#2){\put(#1,#2){\White(-34,-34)
 \dashline{4}(0,0)(-30,-30) \dashline{4}(0,-100)(-32,-39)
 \dashline{4}(-100,0)(-39,-32)}}
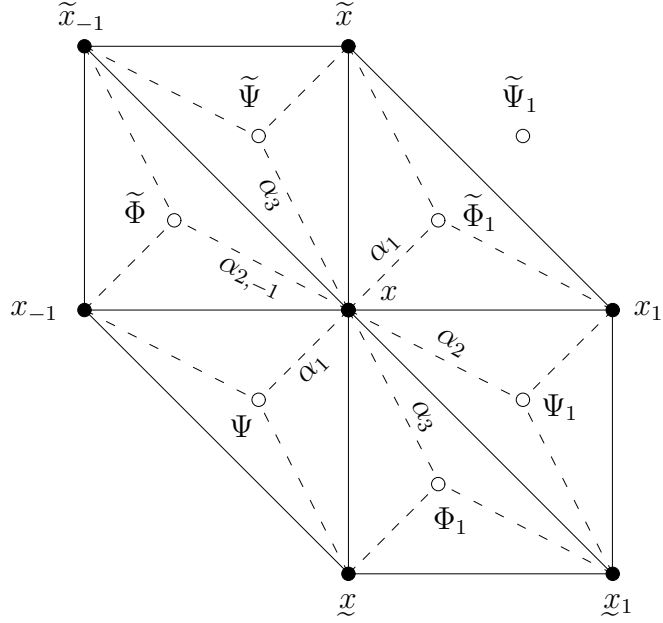
\begin{figure}[htbp]
\begin{center}
\setlength{\unitlength}{0.085em}
\begin{picture}(300,240)(-150,-120)
 \tmpga(-100,0)\tmpga(0,0)\tmpga(0,-100)
 \tmpgb( 100,0)\tmpgb(0,0)\tmpgb(0, 100)
  \path(-100,0)(0,-100) \path(100,0)(100,-100) \path(0,100)(-100,100)
  \put(-110,108){$\ti x_{-1}$} \put(-5,108){$\ti x$}
  \put(-128,-2){$x_{-1}$} \put(12,4){$x$} \put(108,-2){$x_1$}
  \put(-5,-113){$\tu x$} \put(95,-113){$\tu x_1$}
  \put(-45,-48){$\Psi$}
  \put( 32,-83){$\Phi_1$}
  \put( 73,-40){$\Psi_1$}
  \put( 43, 32){$\ti\Phi_1$}
  \put(-42, 77){$\ti\Psi$}
  \put(-85, 33){$\ti\Phi$}
  \put( 58, 77){$\ti\Psi_1$}\White(66,66)
  \put(-20,-28){\rotatebox{ 45}{$\a_1$}}
  \put(  7, 18){\rotatebox{ 45}{$\a_1$}}
  \put( 22,-36){\rotatebox{-60}{$\a_3$}}
  \put(-36, 48){\rotatebox{-60}{$\a_3$}}
  \put(-51, 16){\rotatebox{-25}{$\a_{2,-1}$}}
  \put( 32,-11){\rotatebox{-30}{$\a_2$}}
\end{picture}
\caption{Fields and wave functions on the triangular lattice}
\label{fig:L3}
\end{center}
\end{figure}
Equation of the {\em discrete elliptic Toda system on the
triangular lattice} reads:
\begin{multline}\label{triang Toda}
 f(x,x_1;-\b)+f(x,\ti x;\eps)+f(x,\ti x_{-1};\b_{-1}-\eps)\\
 +f(x,x_{-1};-\b_{-1})+f(x,\tu x;\eps)+f(x,\tu x_1;\b-\eps)=0.
\end{multline}
Using the fact that $f(x,x';-\a)=-f(x,x';\a)$, we put the {\em
Hamiltonian form} of this equation as:
\begin{align}
\label{triang Toda p}
 p&= -f(x,\ti x;\eps)+f(x,x_1;\b)+f(x,x_{-1};\b_{-1})-
 f(x,\ti x_{-1};\b_{-1}-\eps),\\
\label{triang Toda wp}
 \ti p&= f(\ti x,x;\eps)+f(\ti x,x_1;\b-\eps).
\end{align}

\begin{proposition}\label{prop: triang Toda Lax}
The system (\ref{triang Toda p}), (\ref{triang Toda wp}) admits a zero
curvature representation
\begin{equation}\label{dLax}
 \ti LV=V_1L
\end{equation}
with the matrices
\begin{equation}\label{L7}
 L=L(x,p,\b;\lambda) = e^pM(x,\b;\lambda)+e^{-p}M(x,-\b;-\lambda),
\end{equation}
where the matrix $M(x,\b;\lambda)=N(x;\lambda-\b,\lambda)/\s(2\lambda)$ 
is given by
\begin{equation}\label{M}
  M(x,\b;\lambda)=\frac{\s^2(x+\lambda-\b)\s^2(x-\lambda)}{\s(2x)\s(2\lambda)}
   \begin{pmatrix}
    -\wp(x+\lambda-\b) & \wp(x+\lambda-\b)\wp(x-\lambda) \\ -1 & \wp(x-\lambda)
   \end{pmatrix}.
\end{equation}
The matrices $V$ in (\ref{dLax}) are given by
\begin{equation}\label{V}
 V=V(x,x_{-1},\ti x_{-1},\eps;\lambda)
   = e^{\eps G}M(x,\eps;\lambda)+e^{-\eps G}M(x,-\eps;-\lambda),
\end{equation}
where
\begin{equation}\label{G}
\eps G=f(x,x_{-1};\b_{-1})-f(x,\ti x_{-1};\b_{-1}-\eps).
\end{equation}
\end{proposition}
\begin{proof}
Consider the full system of the equations (\ref{Q}) on the dual kagome
lattice and denote the vertices of the white sublattice as shown on the
Fig.~\ref{fig:L3}. The auxiliary linear problem $\Psi_1=L\Psi$ corresponds
to crossing two edges $(x,\tu x)$ and $(x,\tu x_1)$. Adding the three-leg
equations on the quadrilaterals $(x,\Psi,\tu x,\Phi_1)$ and $(x,\Phi_1,\tu
x_1,\Psi_1)$, and using the formula (\ref{triang Toda wp}) we find:
\begin{equation}\label{Mob for Lax L}
 f(x,\Psi;\lambda)-f(x,\Psi_1;\lambda-\b)=
 f(x,\tu x;\eps)+f(x,\tu x_1;\b-\eps)=p.
\end{equation}
Similarly, the auxiliary linear problem $\ti\Psi=V\Psi$ corresponds to
crossing two edges $(x,x_{-1})$ and $(x,\ti x_{-1})$. To find $V$, add
the three-leg equations on two quadrilaterals $(x,\ti\Phi,\ti
x_{-1},\ti\Psi)$ and $(x,\Psi,x_{-1},\ti\Phi)$:
\begin{equation}\label{Mob for Lax V}
 f(x,\Psi;\lambda)-f(x,\ti\Psi;\lambda-\eps)=\eps G,
\end{equation}
with $\eps G$ defined by (\ref{G}). Now Lemma \ref{ZCR matrices} applied
to (\ref{Mob for Lax L}), (\ref{Mob for Lax V}), yields the formulas
(\ref{L7}), (\ref{V}), respectively.
\end{proof}
Of course, in the Hamiltonian framework the main role in the zero
curvature representation belongs to $L$, the transition matrix
along the lattice. For instance, the so-called monodromy matrix
$\prod_k L(x_k,p_k,\b_k;\lambda)$ is a generating function of integrals
of motion. In this context it is important that $L_k$ is local,
i.e. depends on one pair of canonically conjugate variables
$x_k,p_k$ only. As for the matrix $V$, it specifies just one
(discrete) flow of the hierarchy. In the next section we will
present another (continuous) commuting flow.
\section{Elliptic Ruijsenaars--Toda lattice}\label{s:RTL}

The Lagrangian (\ref{triang Toda Lagr}) has a well defined limit
at $\eps\to0$, in the sense that $\cL(X,X+\eps\dot
X)\approx\eps\rL(X,\dot X)$. Alternatively, this limit can be
performed directly in the equations of motion (\ref{triang Toda
p}), (\ref{triang Toda wp}). Taking into account the relations
\[
 \lim_{\eps\to0}f(\ti x,x;\eps)=\log\frac{\dot x+1}{\dot x-1},\quad
 \lim_{\eps\to0}\frac1\eps(f(\ti x,x;\eps)+f(x,\ti x;\eps))=2\z(2x),
\]
one finds, as the limit of the equation (\ref{triang Toda wp}) and
the one obtained by subtracting (\ref{triang Toda p}) from
(\ref{triang Toda wp}):
\begin{align}
\label{rel Toda p}
 p &= \log\frac{\dot x+1}{\dot x-1}+f(x,x_1;\b),\\
\label{rel Toda pt}
 \dot{p} & =  \dot{x}\,\frac{\partial f(x,x_1;\b)}{\partial x}
 +\dot{x}_{-1}\frac{\partial f(x,x_{-1};\b_{-1})}{\partial x_{-1}}
 -\frac{\partial f(x,x_1;\b)}{\partial\beta}
 -\frac{\partial f(x,x_{-1};\b_{-1})}{\partial\beta_{-1}}+2\z(2x).
\end{align}
From here the Newtonian equations of the {\em elliptic
Ruijsenaars--Toda lattice} follow:
\begin{align}
\nonumber
 \frac{\ddot x}{\dot x^2-1} &=
 \dot{x}_1\frac{\partial f(x,x_1;\b)}{\partial x_1}
 -\dot{x}_{-1}\frac{\partial f(x,x_{-1};\b_{-1})}{\partial x_{-1}} \\
\label{ERTL}
 &\qquad +\frac{\partial f(x,x_1;\b)}{\partial\beta}
 +\frac{\partial f(x,x_{-1};\b_{-1})}{\partial\beta_{-1}}-2\z(2x),
\end{align}
where
\begin{align*}
 \frac{\pd f(x,x_1;\b)}{\pd x_1}&=
  \frac12\big(\z(x+x_1+\b)-\z(x-x_1+\b)-\z(x+x_1-\b)+\z(x-x_1-\b)\big),\\
 \frac{\pd f(x,x_1;\b)}{\pd\b}&=
  \frac12\big(\z(x+x_1+\b)+\z(x-x_1+\b)+\z(x+x_1-\b)+\z(x-x_1-\b)\big).
\end{align*}
 A remarkable feature of the zero curvature representation of
Proposition \ref{prop: triang Toda Lax} is that the matrix $L$ in
(\ref{L7}) does not depend on $\eps$. Therefore the flow (\ref{rel
Toda p}), (\ref{rel Toda pt}) shares the matrix $L$ with the map
(\ref{triang Toda p}), (\ref{triang Toda wp}). In other words, the
latter is a B\"acklund transformation of the former.

\begin{proposition}
Elliptic RTL (\ref{rel Toda p}), (\ref{rel Toda pt}) admits a zero
curvature representation
\begin{equation}\label{Lax}
 \dot L=A_1L-LA
\end{equation}
with $L$ from (\ref{L7}) and  $A=A(x,x_{-1},\dot x_{-1};\lambda)
=dV/d\eps|_{\eps=0}$, where $V$ is given in (\ref{V}).
\end{proposition}

By the usual substitution $u=\wp(x)$, the elliptic RTL (\ref{ERTL}) is 
brought into a rational form (which appeared first in \cite{AS97a}):
\begin{equation}\label{ERTL rat}
 \frac{2\ddot u-r'(u)}{\dot u^2-r(u)}
  =-\frac{\dot u_1}{h(u,u_1;\b)}+\frac{\dot u_{-1}}{h(u,u_{-1};\b_{-1})}
  +\frac{\pd}{\pd u}\log\big(h(u,u_1;\b)h(u,u_{-1};\b_{-1})\big),
\end{equation}
where $h$ is a biquadratic polynomial with discriminant $r(u)$ defined in
Appendix \ref{Appendix B}. By the derivation of eq. (\ref{ERTL rat}) from 
(\ref{ERTL}) the following alternative expression is used:
\begin{align*}
 2\frac{\partial f(x,x_1;\b)}{\pd x_1} & =  
 \frac{\s(2\b)\s(2x)\s(2x_1)}
 {\s(x+x_1+\b)\s(x-x_1+\b)\s(x+x_1-\b)\s(x-x_1-\b)}\\
 & =  -\frac{\wp'(x)\wp'(x_1)}{h(u,u_1;\b)}.
\end{align*}
It is obtained with the help of (\ref{4zetas}), (\ref{H mult}), (\ref{h}).
As a consequence, one finds:
\begin{align*}
 2\frac{\partial f(x,x_1;\b)}{\pd\b} &= -\frac{\pd}{\pd x}\log 
\frac{\partial f(x,x_1;\b)}{\pd x_1}+2\z(2x)\\
   &= \wp'(x)\frac{\pd}{\pd u}\log h(u,u_1;\b)-
    \frac{\wp''(x)}{\wp'(x)}+2\z(2x),
\end{align*}
which is the second expression used in the derivation of eq. (\ref{ERTL rat}).

\section{Triangular lattice in $\mathbf \Integer^3$}
\label{s:uv}

\begin{proposition}
Consider a solution $x:V(T)\mapsto\Complex$ of the discrete Toda system
(\ref{triang Toda}) on the triangular lattice $T$. Define a function
$y:V(T)\mapsto\Complex$ by equations
\begin{equation}\label{dYam x}
 F(x,\tu x;\eps)=F(x,y;\b)/F(x,\tu x_1;\b-\eps).
\end{equation}
Then also the following equations hold:
\begin{equation}\label{dYam y}
 F(y,\ti y;\eps)=F(y,x;\b)/F(y,\ti y_{-1};\b-\eps).
\end{equation}
Moreover, $y$ is also a solution of the discrete Toda system
(\ref{triang Toda}).
\end{proposition}

\setcircle{6}
\begin{figure}[htb]
\begin{center}
\setlength{\unitlength}{0.06em}
\begin{picture}(240,200)(-120,-100)
 \texture{c 0 0 0 c 0 0 0
          c 0 0 0 c 0 0 0
          c 0 0 0 c 0 0 0
          c 0 0 0 c 0 0 0}
 \shade\path(100,0)(-100,0)(-50,87)(50,-87)(-50,-87)(50,87)(100,0)
 \path(100,0)(50,-87)\path(-100,0)(-50,-87)\path(-50,87)(50,87)
     \Black(-50,87)\Black(50,87)
     \Black(-50,29)\Black(50,29)
 \Black(-100,0)\Black(0,0)\Black(100,0)
               \Black(0,-58)
     \Black(-50,-87)\Black(50,-87)
 \dashline{5}(-65,29)( 65,29)
 \dashline{5}(-57,42)( 8,-72)
 \dashline{5}( 57,42)(-8,-72)
 \put(8,-60){$y$} \put(55,13){$\ti y$}  \put(-67,13){$\ti y_{-1}$}
 \put(-73, 94){$\ti x_{-1}$} \put(55,94){$\ti x$}
 \put(-127,-2){$x_{-1}$} \put(10,5){$x$} \put(107,-2){$x_1$}
 \put(-65,-95){$\tu x$} \put(55,-95){$\tu x_1$}
\end{picture}
\caption{Map $x\mapsto y$}
\label{fig:xy}
\end{center}
\end{figure}

The geometric description of equation (\ref{dYam x}) is the following: the
variables $y$ should be thought of as defined on the vertices of the new
copy of triangular lattice which correspond to the half of the faces of
the original one, namely those shaded on the Fig.~\ref{fig:xy} (so, $y$
sits on vertex dual to the triangle $(x,\tu x,\tu x_1)$). Then  equation
(\ref{dYam y}) reflects the fact that the variables $x$ are attached to
the half of the faces of the $y$-lattice. Moreover, equations (\ref{dYam
x}), (\ref{dYam y}) are nothing but the three-leg forms (\ref{3leg sym})
of the basic equation \ref{Q4} for the tetrahedra $(x,\tu x,\tu x_1,y)$
and $(y,\ti y,\ti y_{-1},x)$. In terms of $u=\wp(x)$, $v=\wp(y)$, the
equations  (\ref{dYam x}), (\ref{dYam y}) read:
\[
 Q(u,\tu u,\tu u_1,v;\eps,\b)=0,\quad
 Q(v,\ti v,\ti v_{-1},u;\eps,\b)=0.
\]

\begin{figure}[htb]
\begin{center}
\setlength{\unitlength}{0.1em}
\begin{picture}(240,200)(-20,-40)
 \multiput(0,0)(0,60){3}{
  \multiput(0,0)(40,20){3}{\dottedline{2}(0,0)(120,-40)}
  \multiput(0,0)(60,-20){3}{\dottedline{2}(0,0)(80,40)}}
 \multiput(0,0)(40,20){3}{
  \multiput(0,0)(60,-20){3}{\dottedline{2}(0,0)(0,120)}}

 \Black(  0, 60)\put(-20,58){$x_{-1}$}
 \Black(100, 60)\put(95,68){$x$}
 \Black(200, 60)\put(205,58){$x_1$}
 \Black( 60,-20)\put( 54,-30){$\tu x$}
 \Black(140,140)\put(140,146){$\ti x$}
 \Black( 40,140)\put( 22,145){$\ti x_{-1}$}
 \Black(160,-20)\put(159,-30){$\tu x_1$}

 \Black(160,100)\put(163,94){$\ti y$}
 \Black( 60,100)\put( 43,95){$\ti y_{-1}$}
 \Black(120, 20)\put(112,13){$y$}

 \White( 60, 40)\put( 50,30){$\Psi$}
 \White(160, 40)\put(163,30){$\Psi_1$}
 \White(100,120)\put(96,125){$\ti\Psi$}
 \put(64,68){$\a_{2,-1}$} \put(103,102){$\a_3$}
 \put(66,50){$\a_1$} \put(125,43){$\a_2$}
 \path(0,60)(200,60)\path(40,140)(160,-20)\path(60,-20)(140,140)
\end{picture}
\caption{Embedding of the triangular lattice into the cubic one}
\label{fig:star6}
\end{center}
\end{figure}
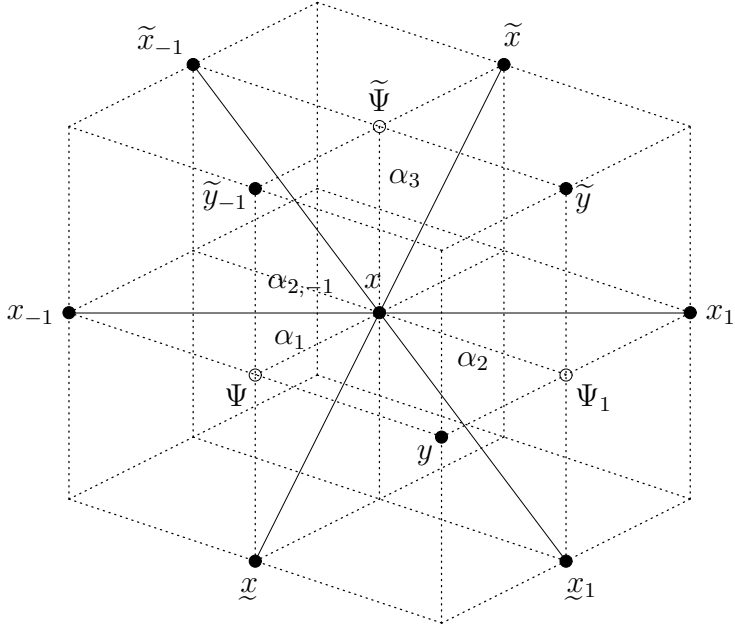

\begin{proof}
Consider the system on the cubic lattice
$\Integer^3=\{(i_1,i_2,i_3)\}$ consisting of equation \ref{Q4} on
each square face, with the parameters $\a_m$ attached to the edges
of the direction $i_m$. This is possible due to the 3D consistency
of equation \ref{Q4} established in Proposition \ref{prop:3d}.
Vertices of the $x$- and $y$-lattices can be interpreted as the
points of the cubic lattice lying in the planes $i_1+i_2+i_3=0$
and $i_1+i_2+i_3=2$, respectively, see Fig.~\ref{fig:star6}. Due
to the three-leg property, both fields $x$ and $y$ satisfy the
discrete Toda system (\ref{triang Toda}). Equations  (\ref{dYam
x}), (\ref{dYam y}) are nothing but the tetrahedron equations
(\ref{tetr}) for the corresponding cubes.

Conversely, it is easy to see that, given a solution $x$ of (\ref{triang
Toda}), one reconstructs uniquely the black sublattice of the cubic
lattice, and then  an arbitrary value at one white vertex determines the
white sublattice uniquely.
\end{proof}

\setcircle{4}

So, equations (\ref{dYam x}), (\ref{dYam y}) can be considered as
a sort of B\"acklund transformation $x\mapsto y$ of the discrete
Toda system (\ref{triang Toda}). On the other  hand, they can be
considered as a discrete time system (mapping) $(\tu
X,Y)\mapsto(X,\ti Y)$, where $X=(x_k)_{k\in\Integer}$,
$Y=(y_k)_{k\in\Integer}$.  In this interpretation, this mapping
serves as discretization of the Shabat-Yamilov lattice considered
in the next section.

\begin{proposition}
The discrete time system (\ref{dYam x}), (\ref{dYam y})
admits a zero curvature representation (\ref{dLax}) with the matrices
\begin{align}
\label{ShYam L}
 L(x,y,\b;\lambda) &= U(u,v;\lambda,\lambda-\b),\\
\label{ShYam V}
 V(x,\ti y_{-1},\eps;\lambda) &= U(u,\ti v_{-1};\lambda,\lambda-\eps),
\end{align}
where $u=\wp(x)$, $v=\wp(y)$.
\end{proposition}
\begin{proof}
The matrices $L$, $V$ correspond, as before, to the transitions
$\Psi\to\Psi_1$ and $\Psi\to\ti\Psi$. As can be seen from
Fig.~\ref{fig:star6}, these are transitions across diagonals of
square faces $(x,\Psi,y,\Psi_1)$ and $(x,\Psi,\ti y_{-1},\ti\Psi)$
of the cubic lattice.
\end{proof}

\section{Shabat-Yamilov lattice}
\label{s:SY}

The continuous limit $\eps\to0$ in equations (\ref{dYam x}),
(\ref{dYam y}) leads to the differential-difference equations
\begin{equation}\label{ShYam}
 \frac{\dot x+1}{\dot x-1}=\frac{F(x,y;\b)}{F(x,x_1;\b)},\quad
 \frac{\dot y-1}{\dot y+1}=\frac{F(y,x;\b)}{F(y,y_{-1};\b)}.
\end{equation}
It should be noticed that elimination of $y$ from these equations
yields ERTL (\ref{ERTL}).

Under the change of variables $u=\wp(x)$, $v=\wp(y)$, the system
(\ref{ShYam}) is brought into a rational form:
\begin{equation}\label{ShYam rat}
 \dot u=\frac{2\rho(u_1,u,v;\b)}{u_1-v},\quad
 \dot v=\frac{2\rho(u,v,v_{-1};\b)}{u-v_{-1}}
\end{equation}
where the polynomial $\rho$ is given in Appendix \ref{Appendix B}.
In the derivation of (\ref{ShYam rat}) from (\ref{ShYam}) one uses the 
formula
\[
 \frac{F(x,y;\b)}{F(x,x_1;\b)}=\frac{\wp(y)-\wp(x+\b)}{\wp(y)-\wp(x-\b)}
 \cdot\frac{\wp(x_1)-\wp(x-\b)}{\wp(x_1)-\wp(x+\b)},
\]
and then takes (\ref{rho}) into account.

The system (\ref{ShYam rat}) was discovered in \cite{SY90}.  Its relation 
to the semi-discretization of the LL equation is discussed in \cite{A00c}. 

\begin{proposition}
The system (\ref{ShYam rat}) admits a zero curvature
representation (\ref{Lax}) with the matrices $L$ from (\ref{ShYam
L}) and
$A=A(u,v_{-1};\lambda)=dU(u,v_{-1};\lambda,\lambda-\eps)/d\eps|_{\eps=0}$.
\end{proposition}

\section{Elliptic Volterra lattice}\label{s:Volt}

\begin{figure}[htb]
\begin{center}
\setlength{\unitlength}{0.1em}
\begin{picture}(300,100)(10,-15)
 \multiput(0,0)(0,60){2}{
  \dottedline{2}(0,0)(240,0)\dottedline{2}(40,20)(280,20)}
 \multiput(0,0)(60,0){5}{\dottedline{2}(0,0)(40,20)(40,80)(0,60)(0,0)}
 \multiput(0,0)(120,0){2}{\path(0,0)(100,20)(120,0)\path(40,80)(60,60)(160,80)}
 \multiput(0,0)(120,0){3}{\Black(0,0)}
 \multiput(40,80)(120,0){3}{\Black(0,0)}
 \multiput(100,20)(120,0){2}{\Black(0,0)}
 \multiput(60,60)(120,0){2}{\Black(0,0)}
 \multiputlist(40,84)(60,0)[bc]{$\ti x_{-1}$,$\ti\Psi$,$\ti x_1$,,$\ti x_3$}
 \multiputlist(4,56)(60,0)[lt]{$\ti\Psi_{-1}$,$\ti x$,$\ti\Psi_1$,$\ti x_2$}
 \multiputlist(37,23)(60,0)[rb]{$\Psi$,$x_1$,,$x_3$}
 \multiputlist(0,-5)(60,0)[tc]{$x$,$\Psi_1$,$x_2$,,$x_4$}
 \White(40,20)\White(60,0)\White(120,60)
 \White(0,60)\White(100,80)
 \put(195,-8){$\a_1=\lambda$} 
 \put(263,5){$\a_2=\lambda-\b$} 
 \put(283,45){$\a_3=\lambda-\eps$}
\end{picture}
\caption{Discrete time Volterra lattice}
\label{fig:volt}
\end{center}
\end{figure}
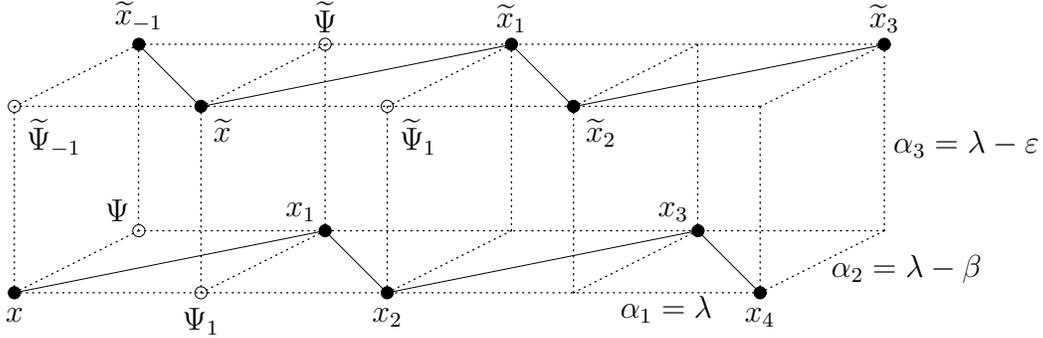

The same system on the cubic lattice that produced in Sect.
\ref{s:uv} the discrete time Shabat-Yamilov lattice serves as the
origin of another important lattice system and its time
discretization. To show this, consider  one row of elementary cubes
and denote the fields variables as on Fig. \ref{fig:volt}. The
parameters  $\a_i$ take the same values (\ref{dual kagome param})
as before. Note that in these notations $\a_2$ (and therefore
$\b$) are site-independent. Then it is easy to see that the fields
$x$ and $\ti x$ satisfy the equation
\begin{equation}\label{dV}
 F(x,\ti x;\eps)=F(x,x_1;\b)/F(x,\ti x_{-1};\b-\eps).
\end{equation}
This equation is nothing but the three-leg form (\ref{3leg sym})
of the basic equation \ref{Q4} for the tetrahedron $(x,x_1,\ti
x,\ti x_{-1})$. It defines an (implicit) map $X\mapsto\ti X$,
where, as usual, $X=(x_k)_{k\in\Integer}$.
\setcircle{4}

\begin{proposition}
The map (\ref{dV}) admits a zero curvature representation (\ref{dLax})
with the matrices
\begin{align*}
 L=L(x_1,x,\b;\lambda) &= U(u_1,u;\lambda,\lambda-\b),\\
 V=V(x_1,\ti x_{-1},\eps;\lambda) &= U(u_1,\ti u_{-1};\lambda,\lambda-\eps).
\end{align*}
\end{proposition}
\begin{proof}
As usual, the matrix $L$ corresponds to the transition
$\Psi\to\Psi_1$, the matrix $V$ corresponds to the transition
$\Psi\to\ti\Psi$. One sees readily from Fig. \ref{fig:volt} that
these are transitions across diagonals of elementary squares.
\end{proof}

In the limit $\eps\to 0$ the equation (\ref{dV}) turns into the
{\em elliptic Volterra lattice}
\begin{equation}\label{eV}
 \frac{\dot{x}-1}{\dot{x}+1}=\frac{F(x,x_1;\b)}{F(x,x_{-1};\b)}.
\end{equation}
In the rational form,
\[
\dot u=\frac{2\rho(u_1,u,u_{-1};\b)}{u_1-u_{-1}},
\]
it is due to Yamilov \cite{Y83}. This lattice admits a zero
curvature representation (\ref{Lax}) with the same matrix $L$ as
in the discrete time model. In particular, the map (\ref{dV}) is a
B\"acklund transformation for the flow (\ref{eV}).

\section{Square lattice, implicit scheme}\label{s:5TLi}

\begin{figure}[htbp]
\begin{center}
\setlength{\unitlength}{0.07em}
\begin{picture}(300,240)(-150,-120)
 \path(-100,0)(-100,100)(100,-100)(100,0)(0,100)(0,-100)(-100,0)
 \Black(-100,100)\Black(0, 100)
 \Black(-100,  0)\Black(0,   0)\Black(100,   0)
                 \Black(0,-100)\Black(100,-100)
 \White(-50,100)\White(50, 100)
 \White(-50,  0)\White(50,   0)
                \White(50,-100)
 \multiput(-100,0)(50,0){3}{
  \multiput(0,0)(50,-100){2}{\dashline{4}(1.5,97)(48.5,3)}}
 \def\tmpg(#1,#2){\multiput(#1,#2)(50,0){2}{\dashline{4}(3.5,0)(46.5,0)}}
 \tmpg(-100,100)\tmpg(-100,0)\tmpg(0,0)\tmpg(0,-100)
 \multiputlist(-100,105)(50,0)[cb]
    {$\ti x_{-1}$,$\ti\Psi$,$\ti x$,$\ti\Psi_1$}
 \multiputlist(-50,5)(100,0)[lb]{$\Psi$,$\Psi_1$}
 \multiputlist(0,-105)(50,0)[ct]{$\tu x$,$\tu\Psi_1$,$\tu x_1$}
 \put(-128,-2){$x_{-1}$}\put(5,5){$x$}\put(108,-2){$x_1$}
 \multiputlist(-25,-5)(50,0)[lt]{$\a_1$,$\a_1$}
 \put(-26,60){$\a_3$} \put(29,-55){$\a_3$}
\end{picture}
\caption{Fields and wave functions on a skew square lattice}
\label{fig:L4i}
\end{center}
\end{figure}
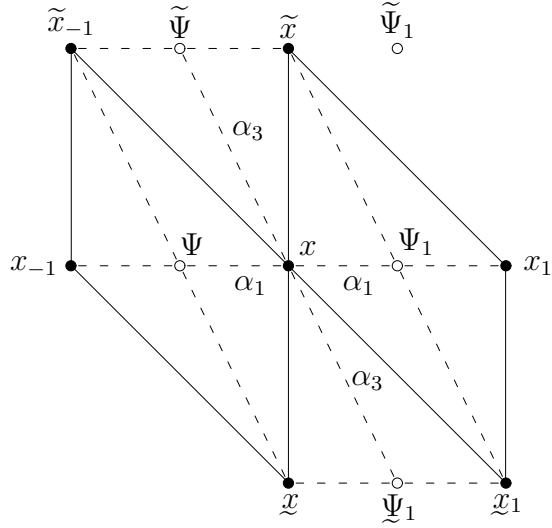

We consider now the skew square lattice with a slicing as on
Fig.~\ref{fig:slice22}, obtained from the triangular one by
erasing the horizontal edges. This can be achieved by setting
$\a_1=\a_2$, so that for our particular setting (\ref{dual kagome
param}) all parameters $\b_k$ of the Toda lattice vanish, and the
parameters of the corresponding quad-graph  (which is a finer
square lattice shown by dashes on Fig.~\ref{fig:L4i}) are
$\a_1=\lambda$, $\a_3=\lambda-\eps$.
\setcircle{5}

All results on the discrete Toda system on the skew square lattice
are thus obtained from those of Sect.\ref{s:7TL} by setting
$\b=0$. The discrete Lagrange function obtained by summing up the
elementary Lagrangians along all edges of a time slice, is equal
to
\[
 \cL(\tu X,X)=\sum_k\bigl(
   \Lambda(x_k,\tu x_k;\eps)
  +\Lambda(x_k,\tu x_{k+1};-\eps)\bigr).
\]
The equations of the Toda system read:
\begin{equation}\label{dETL impl}
 f(x,\ti x;\eps)+f(x,\ti x_{-1};-\eps)+f(x,\tu x;\eps)+f(x,\tu x_1;-\eps)=0.
\end{equation}
The Hamiltonian form of these equations is
\begin{equation}\label{skew square Toda}
 p= -f(x,\ti x;\eps)+f(x,\ti x_{-1};\eps), \quad
 \ti p=  f(\ti x,x;\eps)-f(\ti x,x_1;\eps).
\end{equation}

\begin{proposition}\label{prop: skew square Toda Lax}
The system (\ref{skew square Toda}) admits a zero curvature representation
(\ref{dLax}) with the matrices
\begin{gather}
\label{L5}
  L=L(x,p;\lambda) = e^pM(x,0;\lambda)+e^{-p}M(x,0;-\lambda), \\
\label{V5}
  V=V(x,\ti x_{-1},\eps;\lambda) = U(u,\ti u_{-1};\lambda,\lambda-\eps).
\end{gather}
\end{proposition}

Rewriting (\ref{dETL impl}) in the multiplicative form,
\[
 \frac{F(x,\tu x_1;\eps)}{F(x,\tu x;\eps)}=
 \frac{F(x,\ti x;\eps)}{F(x,\ti x_{-1};\eps)},
\]
and then performing transformations like those mentioned at the
end of Sect. \ref{s:SY}, we put the discrete Toda system in the
rational form:
\[
 \frac{\rho(\tu u_1,u,\tu u;\eps)}{\tu u_1-\tu u}=
 \frac{\rho(\ti u,u,\ti u_{-1};\eps)}{\ti u-\ti u_{-1}}.
\]
In this form it was given in \cite{A00c}.

\section{Elliptic Toda lattice}\label{s:ETL}

Performing the limit $\eps\to 0$ in eqs. (\ref{skew square Toda})
(or $\b\to 0$ in eqs. (\ref{rel Toda p}), (\ref{rel Toda pt})), we find:
\begin{align}
\label{ell Toda p}
 p &= \log\frac{\dot x+1}{\dot x-1},\\
\label{ell Toda pt}
 \dot p &=
 -\z(x+x_1)+\z(x_1-x)-\z(x+x_{-1})-\z(x-x_{-1})+2\z(2x).
\end{align}
From (\ref{ell Toda p}), (\ref{ell Toda pt}) there follow the
Newtonian equations of motion:
\[
 \frac{\ddot x}{\dot x^2-1}=
  \z(x_1+x)-\z(x_1-x)+\z(x+x_{-1})+\z(x-x_{-1})-2\z(2x).
\]
This is the {\em elliptic Toda lattice} as given in \cite{K}. The
rational form of this equation (with $u=\wp(x)$, as usual) was
discovered earlier in \cite{SY90,Y93} and reads:
\[
 \frac{\ddot u-r'(u)/2}{\dot u^2-r(u)}=\frac1{u-u_1}+\frac1{u-u_{-1}}.
\]

The matrix $L$ the zero curvature representation of Proposition
\ref{prop: skew square Toda Lax} does not depend on $\eps$. As a
consequence, this matrix serves also for the continuous time
system.
\begin{proposition}\label{prop: ell Toda Lax}
The system (\ref{ell Toda p}), (\ref{ell Toda pt}) admits a zero
curvature representation (\ref{Lax}) with the matrices $L$ from
(\ref{L5}) and
$A=A(u,u_{-1};\lambda)=dU(u,u_{-1};\lambda,\lambda-\eps)/d\eps|_{\eps=0}$.
\end{proposition}
An alternative zero curvature representation was given in \cite{K}
(with a less local matrix $L$ depending on $x,p$ and $x_1$). The
relation of both representation remains unclear.
\section{Square lattice, explicit scheme}\label{s:5TLe}

\begin{figure}[htbp]
\begin{center}
\setlength{\unitlength}{0.06em}
\begin{picture}(300,240)(-150,-120)
 \multiput(-100,-100)(100,0){3}{\path(0,0)(0,200)
    \Black(0,0)\Black(0,100)\Black(0,200)}
 \multiput(-100,-100)(0,100,0){3}{\path(0,0)(200,0)}
 \multiput(-100,-100)(100,0){2}{\multiput(0,0)(0,100){2}{\White(50,50)
 \dashline{4}(0,0)(46,46) \dashline{4}(100,100)(54,54)
 \dashline{4}(100,0)(54,46)\dashline{4}(0,100)(46,54)}}
 \multiputlist(-100, 105)(100,0)[cb]{$\ti x_{-1}$,$\ti x$,$\ti x_1$}
 \put(-128,-2){$x_{-1}$}\put(12,4){$x$}\put(108,-2){$x_1$}
 \multiputlist(-100,-105)(100,0)[ct]{$\tu x_{-1}$,$\tu x$,$\tu x_1$}
 \multiputlist(-50, 56)(100,0)[cb]{$\ti\Psi$,$\ti\Psi_1$}
 \multiputlist(-50,-44)(100,0)[cb]{$\Psi$,$\Psi_1$}
 \put(-25,-33){$\a_1$}\put(-25,28){$\a_2$}
\end{picture}
\caption{Fields and wave functions on a square lattice}
\label{fig:L4e}
\end{center}
\end{figure}
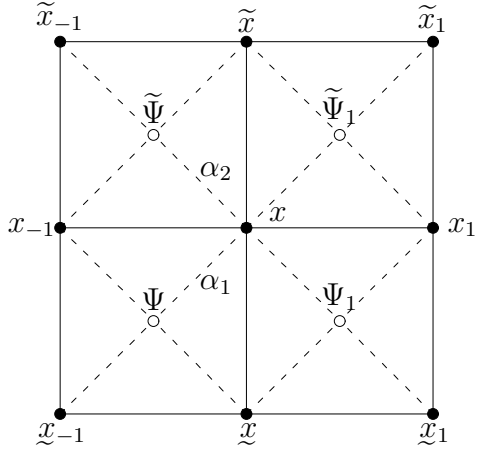

Last, we consider the square lattice with a slicing as on
Fig.~\ref{fig:slice13}, obtained from the triangular one by
erasing the diagonal edges. This can be achieved by setting
$\a_2=\a_3$. For our particular setting (\ref{dual kagome param})
this means that all parameters of the Toda lattice are equal:
$\b_k=\eps$. The parameters of the corresponding quad-graph (finer
square lattice shown by dashes on Fig.~\ref{fig:L4e}) are
$\a_1=\lambda$, $\a_2=\lambda-\eps$.
\setcircle{6}

Correspondingly, the results on the discrete Toda system on the
square lattice are obtained from the results of Sect. \ref{s:7TL}
by putting $\b=\eps$. The discrete Lagrange function is equal to
\[
 \cL(\tu X,X)=\sum_k\bigl(
   \Lambda(x_k,\tu x_k;\eps)
  +\Lambda(\tu x_k,\tu x_{k+1};-\eps)\bigr).
\]
The equations of the discrete Toda system read:
\begin{equation}\label{dETL expl}
  f(x,x_1;-\eps)+f(x,\ti x;\eps)+f(x,x_{-1};-\eps)+f(x,\tu x;\eps)=0,
\end{equation}
and their Hamiltonian form is
\begin{equation}\label{square Toda}
 p= -f(x,\ti x;\eps)+f(x,x_1;\eps)+f(x,x_{-1};\eps),\quad
 \ti p = f(\ti x,x;\eps).
\end{equation}
The rational form of (\ref{dETL expl}) reads:
\[
 \frac{\rho(u_1,u,\tu u;\eps)}{u_1-\tu u}=
 \frac{\rho(\ti u,u,u_{-1};\eps)}{\ti u-u_{-1}}.
\]
The zero curvature representation is of the form (\ref{dLax}), with
\begin{gather*}
  L=L(x,p,\lambda)=e^pM(x,\eps;\lambda)+e^{-p}M(x,-\eps;-\lambda), \\
  V=V(x,x_{-1};\lambda)=U(u,u_{-1};\lambda,\lambda-\eps).
\end{gather*}
Notice also that, in virtue of (\ref{square Toda}), $p=f(x,\tu
x;\eps)$, hence $L=U(u,\tu u;\lambda,\lambda-\eps)$.
\medskip

It is important to mention that both maps (\ref{skew square Toda})
and (\ref{square Toda}) approximate in the continuous limit
$\eps\to 0$ the elliptic Toda lattice (\ref{ell Toda p}),
(\ref{ell Toda pt}). The first (implicit) map is a B\"acklund
transformation of ETL itself (it shares the integrals of motion
and commutes with ETL). However, the second (explicit) map is a
B\"acklund transformation of ERTL with the parameters $\b=\eps$.

\appendix \numberwithin{equation}{section}
\section{List Q of integrable equations on quad-graphs}
\label{Appendix A}

The first three items of the list \ref{Q1}--\ref{Q4} are
\begin{align}
\taQ1     & a(u_0-u_3)(u_1-u_2)-b(u_0-u_1)(u_3-u_2)+\d^2ab(a-b)=0, \\
\nonumber & a(u_0-u_3)(u_1-u_2)-b(u_0-u_1)(u_3-u_2)\\
\taQ2     & \qquad +ab(a-b)(u_0+u_1+u_2+u_3) -ab(a-b)(a^2-ab+b^2)=0,\\
\nonumber & b(a^2-1)(u_0u_1+u_2u_3)- a(b^2-1)(u_0u_3+u_1u_2) \\
\taQ3     & \qquad +(b^2-a^2)(u_0u_2+u_1u_3)-
            \d^2(a^2-b^2)(a^2-1)(b^2-1)/(4ab)=0.
\end{align}
The additive three-leg forms are collected in the table below. 
(In all cases except for (\ref{Q1})$_{\d=0}$ it is actually more convenient
to deal with the multiplicative three-leg form (\ref{3leg mult}).)
\begin{table}[ht]
\def\vv{\vrule height1.5em width0em depth1.6em}
\[
\begin{array}{|l|c|c|c|}
\hline & f(x_0,x_1;\a) & u=\phi(x) & a=\rho(\a) \Big.\\
\hline
(\ref{Q1})_{\d=0} &
  \dfrac{\a}{x_0-x_1}
  & x & \a \vv\\
(\ref{Q1})_{\d=1} &
  \dfrac{1}{2}\log\dfrac{x_0-x_1+\a}{x_0-x_1-\a}
  & x & \a \vv\\
(\ref{Q3})_{\d=0} &
  \dfrac{1}{2}\log\dfrac{\sinh(x_0-x_1+\a)}{\sinh(x_0-x_1-\a)}
  & \exp2x & \exp 2\a \vv\\
(\ref{Q2}) &
 \dfrac{1}{2}\log\dfrac{(x_0+x_1+\a)(x_0-x_1+\a)}{(x_0+x_1-\a)(x_0-x_1-\a)}
  & x^2 & \a \vv\\
(\ref{Q3})_{\d=1} &
 \dfrac{1}{2}\log\dfrac{\sinh(x_0+x_1+\a)\sinh(x_0-x_1+\a)}
                       {\sinh(x_0+x_1-\a)\sinh(x_0-x_1-\a)}
  & \cosh 2x & \exp 2\a \vv\\
(\ref{Q4}) &
 \dfrac{1}{2}\log\dfrac{\s(x_0+x_1+\a)\s(x_0-x_1+\a)}
                       {\s(x_0+x_1-\a)\s(x_0-x_1-\a)}
  & \wp(x) & \wp(\a) \vv\\
 \hline
\end{array}
\]
\end{table}

According to the classification routine developed in \cite{ABS},
the main characteristic of an integrable equation (\ref{Q}) is a
certain polynomial $r(u)$. For the equations \ref{Q1}--\ref{Q4} this
polynomial is equal to
\[
 1,\quad u,\quad u^2-\d^2,\quad 4u^3-g_2u-g_3,
\]
correspondingly. Using this information, one easily finds the
following sequence of degenerations:
\[
 \begin{array}{ccccccc}
  && (\ref{Q3})_{\d=1} &\longrightarrow & (\ref{Q3})_{\d=0} &&\\
  & \nearrow &&&&& \\  
  (\ref{Q4}) && \big\downarrow  && \big\downarrow &&  \\  
  & \searrow &&&&& \\  
  && (\ref{Q2}) &\longrightarrow & (\ref{Q1})_{\d=1} 
                &\longrightarrow & (\ref{Q1})_{\d=0}
 \end{array}
\]
where
\begin{alignat*}{2}
 &(\ref{Q4})\to(\ref{Q3})_{\d=1}:&&
   g_2=\frac43,\quad g_3=-\frac8{27}, \\
 &&& u\to\frac{2}{u-1}+\frac13,\quad
     a\to\frac{4a}{(a-1)^2}+\frac13,\quad
     A\to -\frac{8a(a+1)}{(a-1)^3}, \\
 &(\ref{Q4})\to(\ref{Q2}):&&
   g_2=g_3=0,\quad u\to 1/u,\quad a\to a^{-2},\quad A\to-2a^{-3}, \\
 &(\ref{Q3})_{\d=1}\to(\ref{Q2}):\quad &&
   u\to 1+2\epsilon^2u,\quad a\to1+2\epsilon a,\quad \epsilon\to0, \\
 &(\ref{Q3})_{\d=1}\to(\ref{Q3})_{\d=0}:\quad &&
   u\to u/\epsilon,\quad \epsilon\to0, \\
 &(\ref{Q2})\to(\ref{Q1})_{\d=1}:&&
   a\to\epsilon a,\quad u\to\frac14+\epsilon u,\quad \epsilon\to0, \\
 &(\ref{Q3})_{\d=0}\to(\ref{Q1})_{\d=1}:\quad &&
   u\to 1+2\epsilon u,\quad a\to1+2\epsilon a,\quad \epsilon\to0, \\
 &(\ref{Q1})_{\d=1}\to(\ref{Q1})_{\d=0}:\quad &&
   a\to\epsilon a,\quad \epsilon\to0.
\end{alignat*}

\section{Some formulas with elliptic functions}
\label{Appendix B}

Some of the most useful formulas for the Weierstrass elliptic functions
are:
\begin{align}\label{4sigmas}
 \s(x+\a)\s(x-\a)\s(\b+\g)\s(\b-\g) &=
 \s(x+\b)\s(x-\b)\s(\a+\g)\s(\a-\g)\nonumber\\
   &  \quad -\s(x+\g)\s(x-\g)\s(\a+\b)\s(\a-\b),
\end{align}
\begin{equation}\label{4zetas}
 \z(x)+\z(y)+\z(z)-\z(x+y+z)=\frac{\s(x+y)\s(y+z)\s(z+x)}
 {\s(x)\s(y)\s(z)\s(x+y+z)},
\end{equation}
\begin{equation}\label{det to sigma}
 \frac12\left|\begin{array}{ccc}
  1 & \wp(x) & \wp'(x) \\ 
  1 & \wp(y) & \wp'(y) \\ 
  1 & \wp(z) & \wp'(z)
 \end{array}\right|
  =\frac{\s(x+y+z)\s(x-y)\s(y-z)\s(z-x)}{\s^3(x)\s^3(y)\s^3(z)},
\end{equation}
\begin{equation}\label{p-p}
 \wp(x)-\wp(y)=-\frac{\s(x+y)\s(x-y)}{\s^2(x)\s^2(y)}.
\end{equation}
A fundamental object related to the elliptic curve $\cE$ is the polynomial
\begin{equation}\label{H}
 H(u,v,w)=(uv+vw+wu+g_2/4)^2-(u+v+w)(4uvw-g_3).
\end{equation}
Indeed, the equation
$H(\wp(x),\wp(y),\wp(x\pm y))=0$ is the addition theorem for the
$\wp$-function. This follows from the formula
\begin{equation}\label{H mult}
 H(\wp(x),\wp(y),\wp(z))=
 -\frac{\s(x+y+z)\s(-x+y+z)\s(x-y+z)\s(x+y-z)}{\s^4(x)\s^4(y)\s^4(z)}.
\end{equation}
There holds the identity $H_v^2-2HH_{vv}=r(u)r(w)$. This gives rise to a
one-parameter family of symmetric biquadratics $h(u,v;\b)$
with the property $h_v^2-2hh_{vv}=r(u)$:
\begin{equation}\label{h}
 h(u,v;\b)=H(u,v,b)/\sqrt{r(b)}=H(u,v,\wp(\b))/\wp'(\b).
\end{equation}
Recall \cite{ABS} that the polynomial $h$ is related to the basic
polynomial $Q$ from eq. (\ref{Q}) by the formula
$Q_{u_2}Q_{u_3}-QQ_{u_2u_3}=k(\a,\b)h(u_0,u_1;\a)$.

To an arbitrary quadratic polynomial $h(u)=c_2u^2+2c_1u+c_0$ there
corresponds the symmetric affine-linear polynomial $\rho(u,w)$ with the
property $h(u)=\rho(u,u)$, namely $\rho=c_2uw+c_1(u+w)+c_0$. In a more
invariant fashion, $2\rho=2h-(u-w)h_u$. For the polynomial $h=h(u,v;\b)$
from (\ref{h}) this polynomial $\rho$ is given by:
\begin{equation}\label{rho}
\rho(u,v,w;\b)=
\frac{(\wp(y)-\wp(\b))^2}{\wp'(\b)}\Bigl(
uw-(\wp(y+\b)+\wp(y-\b))\frac{u+w}2+\wp(y+\b)\wp(y-\b)\Bigr),
\end{equation}
where, as usual, $u=\wp(x)$, $v=\wp(y)$ and $w=\wp(z)$. This formula follows
easily from 
\[
 h(u,v;\b)=
 \frac{(\wp(y)-\wp(\b))^2}{\wp'(\b)}(u-\wp(y+\b))(u-\wp(y-\b)),
\]
which, in turn, is a direct consequence of (\ref{H mult}), (\ref{h}).


\end{document}